\documentclass{LT23auth}
\usepackage{graphicx}

\begin{document}

\begin{frontmatter}

\title{Ground state of antiferromagnetic Heisenberg two-leg ladder\\
in terms of the valence-bond solid picture}

\author[address1]{Munehisa Matsumoto \thanksref{thank1}},
\author[address1,address2]{Synge Todo \thanksref{thank2}},
\author[address3]{Masaaki Nakamura},
\author[address1]{Chitoshi Yasuda \thanksref{thank3}},
\author[address1]{Hajime Takayama}

\address[address1]{Institute for Solid State Physics, University of Tokyo,
Kashiwa 277-8581, Japan}

\address[address2]{Theoretische Physik, Eidgen\"ossische Technische
Hochschule, CH-8093 Z\"urich, Switzerland}

\address[address3]{Department of Applied Physics,
Faculty of Science, Science University of Tokyo,
Tokyo 162-8601, Japan}

\thanks[thank1]{ E-mail: matumoto@issp.u-tokyo.ac.jp}
\thanks[thank2]{ Present address: Department of Applied Physics,
University of Tokyo, Tokyo 113-8656, Japan}
\thanks[thank3]{ Present address: Computational Materials Science
Center, National Institute for Materials Science, Tsukuba 305-0047, Japan}

\begin{abstract}
We have proposed the plaquette-singlet solid (PSS) ground state for the
spin-$1$ antiferromagnetic Heisenberg two-leg ladder.  Based on the PSS
picture, we discuss the correspondence of the PSS state
to the valence-bond solid (VBS) state of the ground state
of spin-$2$ chain by introducing an
appropriate composite spin picture.
When the bond alternation is introduced, there
occur quantum phase transitions and each phase can be identified with
that in the dimerized spin-$2$ chain.  Furthermore, we argue that the
VBS picture of spin-$2S$ chain can be applied to the
ground state of spin-$S$ two-leg ladder.
\end{abstract}

%
%
\end{frontmatter}


Ground state of one-dimensional quantum spin systems in general has no
long-range order due to strong quantum fluctuations.  A typical example
investigated so far is the celebrated Haldane gap system.  The
well-established valence-bond solid (VBS) picture~\cite{aklt} gives the
intuitive understanding for the ground state of such spin chains. Especially
for the spin-1 antiferromagnetic (AF) Heisenberg chain the hidden AF
order~\cite{denNijs} in the ground state has been discussed. This is an
interesting argument that shows a disordered state realized by quantum
fluctuations is not totally disordered, but has some kind of
topological hidden order.

In Ref.~\cite{todo}, we have discussed the ground state and its hidden
order of the spin-1 AF Heisenberg two-leg ladder, in which the naive VBS
picture breaks down.  The spin-1 ladder has a gapped ground state
irrespective of the ratio between the strength of the rung coupling and
the leg coupling, including the two limits of independent rung dimers
and two decoupled spin-1 Haldane chains.  In the VBS picture for these two
limiting cases, the valence bonds are located on the rungs and legs,
respectively.  Thus the states characterized by the different
valence-bond configurations are in the same phase.  We focused on a
singlet wavefunction of the four $S=1/2$ spins on the plaquette between
the two nearest rungs.  It can be written as a linear combination of
the rung valence bonds and the leg valence bonds.  We have shown that
the ground state of the spin-1 two-leg ladder is described well by the
wave function constructed from the spin-$1/2$ plaquette
singlets~\cite{todo}.  This is called the plaquette-singlet solid (PSS)
picture.

It is interesting to see that there is no quantum phase transition in
the spin-1 AF ladder with the in-phase bond alternation~\cite{todo,mm},
while it exhibits quantum phase transitions in the presence of the
anti-phase bond alternation~\cite{yamamoto,nakamura}.
What does this difference
with respect to the bond alternation pattern suggest?  Furthermore,
the spin-1 ladder with the ferromagnetic rung coupling is
equivalent to the spin-2 chain in the strong coupling limit, which
exhibits quantum phase transitions between $(2,2)$-, $(3,1)$-, and
$(4,0)$-phases in the presence of bond alternation, where the indices
$(m,n)$ denotes the valence-bond configuration with $m$ ($n$) valence
bonds on odd (even) bonds~\cite{oshikawa}. What about the spin-1 ladder
with the weak ferromagnetic rung coupling and bond alternation?
In the following we will discuss these questions in detail.

%
%

\begin{figure}[t]
\begin{center}\leavevmode
\scalebox{0.7}{\includegraphics[width=1.0\linewidth]{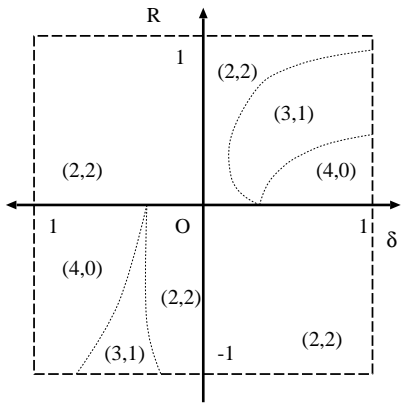}}
\caption{Schematic ground-state phase diagram of the spin-1 ladder.  The
parameters $R$ and $\delta$ denote the strength of rung coupling and the
dimerization, respectively.  The right (left) half shows the anti-phase
(in-phase) bond-alternation case.  There exist quantum critical lines
only in the upper-right and down-left regions.  Each phase in the phase
diagram can be identified with the VBS state for the composite $S=2$
diagonal (rung) spin for $R>0$ ($R<0$).}  \label{phase_diagram}
\end{center}
\end{figure}

The Hamiltonian for the spin-1 two-leg ladder is given by
\begin{equation}
 H=\sum_{i=1,2} \sum_{j} J_{i,j}{\mathbf S}_{i,j}\cdot{\mathbf S}_{i,j+1}
 + K \sum_{j} {\mathbf S}_{1,j}\cdot{\mathbf S}_{2,j},
 \label{ham}
\end{equation}
where ${\mathbf S}_{i,j}$ denotes the $j$th spin on the $i$th leg.  We
consider the cases with in-phase bond alternation $J_{i,j}=J[1+(-1)^j
\delta]$ and with anti-phase bond alternation $J_{i,j}=J[1+(-1)^{i+j}
\delta]$. The schematic ground-state phase diagram, parametrized by the
strength of bond alternation $\delta$ and the rung coupling
$R=K/(J+|K|)$, is shown in Fig.~\ref{phase_diagram}. The phase
boundaries are determined precisely by the quantum Monte Carlo
method~\cite{evertz}.

In the AF rung region ($R>0$) with anti-phase dimerization (right-upper
in Fig.~\ref{phase_diagram}), there are three different phases. They
can be identified as the $(2,2)$-, $(3,1)$-, and $(4,0)$-VBS states for
the spin-2 dimerized chain, respectively. This is done by
making $S=2$ composite spins
from the two $S=1$ spins situated at the diagonal
position in a plaquette between the nearest neighbor rungs.
For example, in the PSS state, where two parallel valence bonds locate on
either the rung edges or the leg edges of a plaquette, diagonal
composite spins always have two valence bonds in-between. Thus the PSS
state is equivalent to the $(2,2)$-VBS state of spin-2 diagonal spin chain.
The {\it non}-existence of the quantum phase transition in the in-phase
dimerization case (left-upper in Fig.~\ref{phase_diagram})
is also well understood by the same procedure, i.e.,
in this whole region the ground state is in
the $(2,2)$-VBS phase for the diagonal composite
$S=2$ spin. The interpretation of
the ferromagnetic rung region ($R<0$) is similar
by making the $S=2$ composite spin from the two $S=1$ spins on a
rung.  On the line $R=-1$, which is equivalent to the spin-2 dimerized
AF chain, there exists two critical points, and they are continued to
the critical point of the spin-1 dimerized chain on the $R=0$ line
separating the $(2,2)$-, $(3,1)$-, and $(4,0)$-VBS states of the spin-2
composite rung spin chain.  We have confirmed that each phase can be
characterized by the spin-2 generalized string order
parameters~\cite{oshikawa,yamamoto2},
and by the ground-state expectation value
of the twist operator~\cite{nakamura}.

The above arguments on the spin-1 ladder is a natural generalization of
that on the spin-1/2 non-dimerized ladder by Nishiyama {\it et
al}~\cite{nishiyama}.  By introducing bond alternation, the similar
phase diagram to that of spin-1 ladder can be drawn.  The generalization
to the cases with larger $S$ is straightforward.

To summarize, for the ground state of spin-$S$ two-leg ladder the naive
VBS picture breaks down. However, by making an appropriate composite
spin-$2S$ chain, we can identify each phase in spin-$S$ ladder as that
appears in spin-$2S$ dimerized chain.  The composite spin is made from the
diagonal spins on the plaquette when the rung coupling is AF, and from
the rung spins when the rung coupling is ferromagnetic.

\end{document}